\documentclass[10pt,conference]{IEEEtran}
\IEEEoverridecommandlockouts
\usepackage{cite}
\usepackage{amsmath,amssymb,amsfonts}
\usepackage{algorithmic}
\usepackage{graphicx}
\usepackage{textcomp}
\usepackage{xcolor}
\def\BibTeX{{\rm B\kern-.05em{\sc i\kern-.025em b}\kern-.08em
T\kern-.1667em\lower.7ex\hbox{E}\kern-.125emX}}

\newcommand{\added}[1]{{\color{black} #1}}

\newcommand{\code}[1]{{\underline{#1}}}

\newcommand{\researchquestion}[1]{{\textbf{RQ#1}}}

\newcommand{\participantQuote}[2]{{
    \parbox{0.95\linewidth}{
        \vspace{2pt}
        \small
        \faQuoteLeft\xspace 
        \emph{#1}" (P#2)
        \vspace{2pt}
    }
}}

\newcommand{\participantQuotes}[4]{{
    \parbox{0.95\linewidth}{
        \vspace{3pt}
        \small
        \faQuoteLeft\xspace 
        \emph{#1}" (P#2)
        \vspace{3pt}
        
        \small
        \faQuoteLeft\xspace 
        \emph{#3}" (P#4)
        \vspace{3pt}
    }
}}

\newcommand{\icon}[1]{{\includegraphics[height=1.5\fontcharht\font`\B]{#1}}\xspace}
\newcommand{\meiicon}{\icon{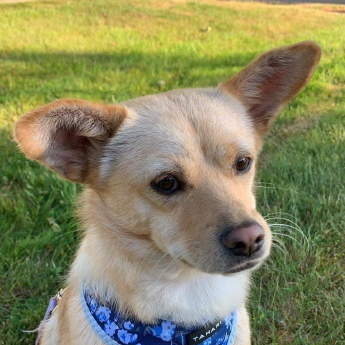}}

\usepackage{booktabs}
\usepackage{fontawesome}
\usepackage{multirow}
\usepackage{url}
\usepackage{xcolor}
\usepackage{xspace}
\usepackage[switch]{lineno}
\begin{document}

\title{A Qualitative Study on the Implementation Design Decisions of Developers}

\makeatletter
\newcommand{\linebreakand}{%
  \end{@IEEEauthorhalign}
  \hfill\mbox{}\par
  \mbox{}\hfill\begin{@IEEEauthorhalign}
}
\makeatother

\author{
  \IEEEauthorblockN{Jenny T. Liang}
  \IEEEauthorblockA{\textit{School of Computer Science} \\
    \textit{Carnegie Mellon University}\\
    Pittsburgh, PA, USA \\
    jtliang@cs.cmu.edu}
  \and
  \IEEEauthorblockN{Maryam Arab}
  \IEEEauthorblockA{\textit{Department of Computer Science} \\
    \textit{George Mason University}\\
    Fairfax, VA, USA \\
    marab@gmu.edu}
  \and
  \IEEEauthorblockN{Minhyuk Ko}
  \IEEEauthorblockA{\textit{Department of Computer Science} \\
    \textit{Virginia Tech}\\
    Blacksburg, VA, USA \\
    minhyukko@vt.edu}
  \linebreakand %
  \IEEEauthorblockN{Amy J. Ko}
  \IEEEauthorblockA{\textit{Information School} \\
    \textit{University of Washington}\\
    Seattle, WA, USA \\
    ajko@uw.edu}
  \and
  \IEEEauthorblockN{Thomas D. LaToza}
  \IEEEauthorblockA{\textit{Department of Computer Science} \\
    \textit{George Mason University}\\
    Fairfax, VA, USA \\
    tlatoza@gmu.edu}
}

\maketitle

\begin{abstract}
Decision-making is a key software engineering skill. Developers constantly make choices throughout the software development process, from requirements to implementation. While prior work has studied developer decision-making, the choices made while choosing what solution to write in code remain understudied. In this mixed-methods study, we examine the phenomenon where developers select one specific way to implement a behavior in code, given many potential alternatives. We call these decisions \emph{implementation design decisions}.
Our mixed-methods study includes 46 survey responses and 14 semi-structured interviews with professional developers about their decision types, considerations, processes, and expertise for implementation design decisions. We find that implementation design decisions, rather than being a natural outcome from higher levels of design, require constant monitoring of higher level design choices, such as requirements and architecture. We also show that developers have a consistent general structure to their implementation decision-making process, but no single process is exactly the same.  
We discuss the implications of our findings on research, education, and practice, including insights on teaching developers how to make implementation design decisions.
\end{abstract}

\begin{IEEEkeywords}
implementation design decisions, software design
\end{IEEEkeywords}

\begin{figure}[ht!]
\centering
\includegraphics[trim=0 0 0 0, clip, width=1.0 \linewidth, keepaspectratio]{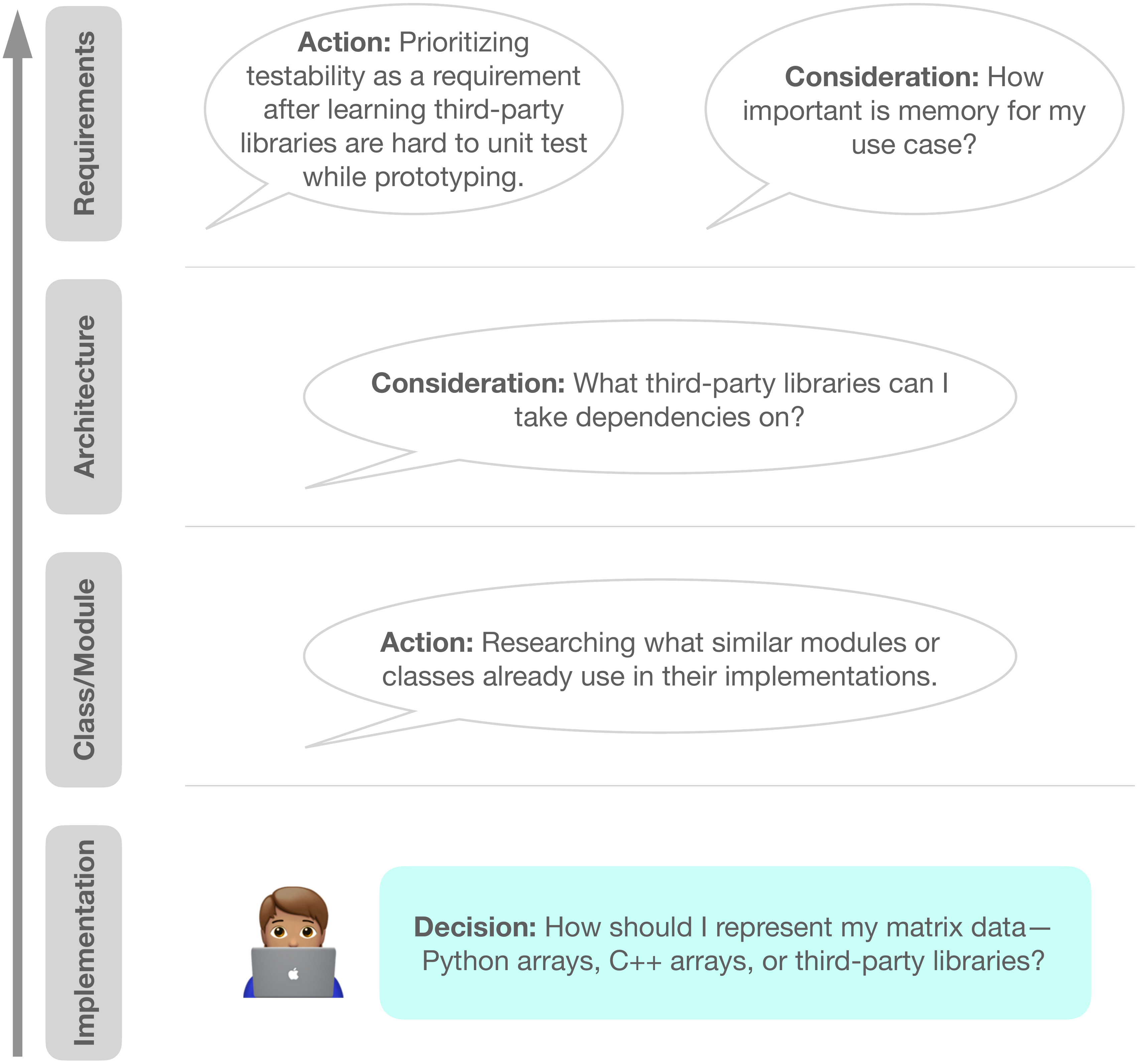} 
\caption{An example of an implementation design decision. Developers consider aspects of software design that are above the implementation (e.g., requirements, architecture, or class and modules) to make these decisions.}
\label{fig:theory}
\end{figure}

\section{Introduction}
Making decisions effectively is a crucial skill for software engineers~\cite{li2015makes}. One reason is because making explicit and rationalized design decisions during the design process improves software design quality~\cite{tang2008design}. 
Developers make these explicit decisions throughout the software design process, from requirements and architecture to implementation. These decisions in turn have downstream effects on the software, such as influencing how easily developers comprehend a codebase~\cite{rajlich2002role} or resulting in systems that are difficult to maintain~\cite{kruchten2012technical,verdecchia2020architectural}. 

At higher levels of software design, developers make explicit decisions about the software architecture by prototyping them at the whiteboard~\cite{cherubini2007let} or documenting them in UML diagrams~\cite{petre2013uml}. Researchers have developed an understanding of how developers make architectural decisions~\cite{van2016decision,tang2017human}, even building several tools to aid this process~\cite{shahbazian2020equal,jansen2007tool}. At the code level, developers also make explicit decisions, such as implementing specific design patterns~\cite{gamma1995design,beck1996industrial}. Yet, the decisions that developers make while choosing what solution to write in code remains understudied. We call these decisions \emph{implementation design decisions}.

Implementation design decisions are when developers select one specific way to implement a behavior, given many potential alternatives. They may choose an implementation that minimizes the time to market in place of producing high-quality, robust software. 
Meanwhile, developers may choose to implement solutions for other reasons: system requirements (e.g., performance), code quality (e.g., readability), or convenience (e.g., ease of implementation). 

Figure~\ref{fig:theory} includes an example of an implementation design decision. A developer is considering three different ways to represent their matrix data in a script---Python arrays, C++ arrays, or using a third-party library. If the developer wanted to minimize runtime, they could use C++ arrays or third-party libraries. If they wanted a simple, easy-to-read solution for their teammates, they could opt for Python arrays.

Understanding implementation design decisions can provide insights on what decisions result in good or bad software designs, which could be taught to novice developers. Furthermore, studying these decisions can elicit a broader set of considerations developers are optimizing for. This could help explain the decisions that are on its face sub-optimal but are in fact necessary (e.g., decisions causing technical debt).

In this work, we study implementation design decisions with a focus on the explicit decisions made by developers, as conscious reasoning in software design improves the design quality~\cite{tang2008design}.
We address the following questions:

\begin{itemize}
    \item \researchquestion{1}: What implementation design decisions do software developers make?
    \item \researchquestion{2}: What considerations do software developers have while making implementation design decisions?
    \item \researchquestion{3}: What process do software developers follow to make implementation design decisions?
    \item \researchquestion{4}: Which types of developer expertise are described in the implementation decision-making process?
\end{itemize}

To answer these questions, we designed a mixed-methods study using surveys and interviews to understand implementation design decisions. Our study had 53 participants who program professionally. 
We find that implementation design decisions demanded careful thought. Overall, they require constant evaluation of higher levels of design and could even exert influence on them (see Figure~\ref{fig:theory}), which corroborates findings from prior work~\cite{ralph2016characteristics}. This is in contrast to other work that characterizes implementation as a natural result from higher level forms of design in a top-down fashion. For example, Perry and Wolf described implementation as a code-level representation satisfying requirements, architecture and design~\cite{perry1992foundations}. \added{Thus, our study supports the theory that problems and solutions in software design co-evolve with one another---as the solution develops, the problem space can update~\cite{van2016decision}.}
We also show that developers have a general structure to their decision-making process, but each developers' process is unique. Finally, we provide implications in how our results can be applied in research, education, and practice.

\section{Related Work}

\subsection{Implementation Design Decisions}
\added{Ralph and Tempero studied the characteristics of decisions made while programming~\cite{ralph2016characteristics}. They elicited common themes in these decisions, such as their considerations and the approaches taken to solve a problem. Our study examines a similar phenomenon as Ralph and Tempero. It extends this work by also investigating the systematic processes developers use to make such decisions, as well as the types of expertise they apply in making them.}

Prior research has also investigated a specific outcome from a particular subset of implementation design decisions in depth---technical debt. These occur when developers choose implementations that bias time to market over software quality~\cite{cunningham1992wycash}. Technical debt comes in several forms, such as architectural, structural, documentation, test, infrastructure, and requirements debts~\cite{kruchten2012technical,verdecchia2020architectural}.
Technical debt has many negative downstream effects, such as reducing developer productivity~\cite{besker2019software} as well as decreasing team morale, causing delays, and lowering code quality~\cite{rios2018common}. Technical debt can be caused for many reasons, including planning and management (e.g., deadlines), development issues (e.g., not adopting good practices), software engineering processes (e.g., a lack of documentation), a lack of knowledge (e.g., lack of experience), human factors (e.g., lack of commitment), organizational factors (e.g., lack of trained professionals), external factors (e.g., pressure), and infrastructure (e.g., unavailable infrastructure)~\cite{verdecchia2020architectural,rios2019supporting}.

Implementation design decisions, such as ones that result in technical debt, are concerned with decisions that developers make when they are about to write an implementation. They consider different ways a behavior may be implemented in code. 
Thus, prior work in this space largely focuses on the far-reaching effects of implementation design decisions, rather than the decisions themselves.

\subsection{Decision-Making in Software Engineering}
Developers make decisions across several types of activities, such as in project planning and verification~\cite{ruhe2002software}. These decisions require developers to make tradeoffs by reasoning about future outcomes~\cite{becker2017intertemporal}. Prior work in software engineering expertise also suggests that knowing how to effectively make decisions is a form of expertise~\cite{baltes2018towards, li2015makes, li2020distinguishes}.

\added{Requirements engineering activities can be framed as decision-making processes at the organizational and individual levels~\cite{aurum2003fundamental}.} While deciding on requirements, software developers make decisions on which requirements to prioritize~\cite{berander2005requirements}. 

In software architecture, developers make decisions to select between candidate architectures~\cite{falessi2011decision,bass2003software} and integrate components into an existing system~\cite{ruhe2002software}. \added{van Vliet et al. argue that software architecture can be viewed as a set of design decisions made by software architects~\cite{van2016decision}.} Additionally, software architecture decisions use both slow, rational and fast, intuitive thinking and can be prone to cognitive biases~\cite{van2016decision,tang2017human}. \added{Zannier et al. found which of the two thinking approaches designers used depended on how structured the problem was~\cite{zannier2007model}.} 

Software developers also make decisions in API design. Stylos and Myers outlined the space of design decisions for APIs, such as design patterns to use or what fields or methods to provide. These decisions were split across the architectural, structural, class, and language levels~\cite{stylos2007mapping}. In order to make proper API design decisions, software practitioners and organizations have outlined several guidelines~\cite{dharani2017web,swift2022api} and online resources~\cite{fowler2022api} on how to design APIs.

Overall, prior work has largely studied developers' decision-making in specific development contexts or at higher levels of design, such as requirements or architecture. Our work diverges from this literature by studying the decisions made while developers are about to write implementations in code.

\subsection{Software Design Practices}
Many studies have examined how software developers do software design. Petre and van der Hoek studied the individual practices of software designers~\cite{petre2016software}, such as involving experts from outside the team to learn domain-specific knowledge. Practitioners have also written resources on writing code with clean software design (e.g.,~\cite{bloch2008effective,martin2009clean}) and architecture (e.g., ~\cite{martin2018clean,fairbanks2010just}) with prescriptions on how to make design decisions effectively, such as following well-known programming principles like SOLID. 
In open-source software, contributors make design decisions on bug reports~\cite{ko2011design} and online discussions on GitHub issues~\cite{brunet2014developers,mahadi2022conclusion}. 

For software design at higher levels of abstraction, developers are often known to perform design activities at the whiteboard, creating visual diagrams to support the design process~\cite{mangano2014supporting,cherubini2007let}. Software developers can informally denote their designs using UML~\cite{petre2013uml} or sketches~\cite{baltes2014sketches}. Prior work has also investigated the process software engineers follow while doing early stage design work. Sharif et al. found that while designing, developers engage in activities which overlap with the traditional software development process, namely requirements, analysis, design, and implementation~\cite{sharif2013identifying}.

Thus, prior research has investigated the practices developers follow to design software. Instead of studying developers' actions, our work contrasts prior literature by studying the decision-making and reasoning of developers.

\section{Study Design}
\label{sec:study-design}
To answer the research questions, we collected data on software developers' implementation design decisions from 53 study participants (Section~\ref{sec:participants}) using surveys (Section~\ref{sec:survey}) and interviews (Section~\ref{sec:interviews}). We then analyzed the data using qualitative coding (Section~\ref{sec:analysis}). The breadth of the survey data and the depth of the interview data corroborated our findings from multiple sources. 

\subsection{Participants}
\label{sec:participants}
To elicit a wide range of developers' insights on implementation design decisions, we recruited developers with diverse programming experiences. 
Our inclusion criteria was developers who contributed their programming expertise to at least one project in a professional setting. 

\subsubsection{Sampling strategy}
We recruited participants with different levels of software engineering experience, technology expertise, job titles, and engineering team sizes. The authors then released a survey (Section~\ref{sec:survey}) on their personal social media accounts for recruitment. Social media posts displayed brief descriptions about implementation design decisions, the estimated time of completion, and a survey link.

We increased sample coverage by also recruiting on Reddit. The first author advertised the survey in a text post on 10 popular developer-centric Reddit communities. As of January 2022, the subreddits had between 6,681 members to 291,170 members. Reddit communities were selected by technology (e.g., r/reactjs, r/php), geographic location (e.g., r/developersIndia), or role (e.g., r/dataengineering). Posts were only published after receiving permission from the community's moderators, as stipulated by the institutional review board. The Reddit recruitment posts introduced  implementation design decisions, provided an estimated time of completion, explained why the subreddit was a good fit, and linked the study. We also relied on snowball sampling by encouraging interview participants to share the study to others. In the survey, participants indicated if they wanted to participate in a follow-up interview.

In total, 60 developers agreed to participate in the study, with 53 who met the inclusion criteria and participated in the study. 46 participants completed the survey and 14 participants completed an interview.

\subsubsection{Demographics}
\added{We report interviewee demographics in Table~\ref{tab:interviewees}.} In our study, participants represented diverse geographic locations, including the Americas ($n=27$), Africa ($n=1$), Asia ($n=7$), Europe ($n=14$), and Oceania ($n=3$). Multiple genders were also represented, such as man ($n=44$), woman ($n=7$), and non-binary ($n=1$). Participants had job titles such as junior, senior, or principal software engineer; Chief Technology Officer; software architect; machine learning engineer; and research engineer. Participants reported contributing between 1 to over 1,000 projects, with a median of 25. Participants worked in companies of varying sizes, whose engineering organization sizes ranged from 1 to 36,000, with a median of 18. Participants reported using a variety of technologies (e.g., Java, MongoDB, Angular, R, Verilog, Jupyter notebooks) and working on diverse problem domains (e.g., deployment infrastructure, IoT, static analysis tools, financial technology, healthcare, online exams).

\begin{table}\centering\small
\caption{\added{Overview of interview participants. ``Software engineer" is abbreviated as ``SWE".}
}\label{tab:interviewees}
\begin{tabular}{p{0.05\linewidth}p{0.24\linewidth}p{0.09\linewidth}p{0.08\linewidth}p{0.08\linewidth}p{0.15\linewidth}}
\toprule
\parbox[c]{\hsize}{\textbf{ID}} & \parbox[c]{\hsize}{\textbf{Job title}} & \parbox[c]{\hsize}{\textbf{\# projects}} & \parbox[c]{\hsize}{\textbf{Org. size}} & \parbox[c]{\hsize}{\textbf{Gender}} & \parbox[c]{\hsize}{\textbf{Location}} \\\midrule
\parbox[c]{\hsize}{P4} & \parbox[c]{\hsize}{Senior Principal Engineer} & \parbox[c]{\hsize}{50+} & \parbox[c]{\hsize}{1,700} & \parbox[c]{\hsize}{Man} & \parbox[c]{\hsize}{United Kingdom}\\ \midrule
\parbox[c]{\hsize}{P6} & \parbox[c]{\hsize}{Chief Technology Officer} & \parbox[c]{\hsize}{30} & \parbox[c]{\hsize}{16} & \parbox[c]{\hsize}{Man} & \parbox[c]{\hsize}{Germany}\\ \midrule
\parbox[c]{\hsize}{P7} & \parbox[c]{\hsize}{Scientific SWE II} & \parbox[c]{\hsize}{20} & \parbox[c]{\hsize}{200} & \parbox[c]{\hsize}{Woman} & \parbox[c]{\hsize}{United States}\\ \midrule
\parbox[c]{\hsize}{P15} & \parbox[c]{\hsize}{Head of Research} &  \parbox[c]{\hsize}{Dozens} & \parbox[c]{\hsize}{12} & \parbox[c]{\hsize}{Man} & \parbox[c]{\hsize}{Israel} \\ \midrule
\parbox[c]{\hsize}{P18} & \parbox[c]{\hsize}{Principal SWE} & \parbox[c]{\hsize}{50} & \parbox[c]{\hsize}{1,000} & \parbox[c]{\hsize}{Man} & \parbox[c]{\hsize}{United States}\\ \midrule
\parbox[c]{\hsize}{P22} & \parbox[c]{\hsize}{Chief Technology Officer} & \parbox[c]{\hsize}{10} & \parbox[c]{\hsize}{6} & \parbox[c]{\hsize}{Man} & \parbox[c]{\hsize}{Philippines}\\ \midrule
\parbox[c]{\hsize}{P28} & \parbox[c]{\hsize}{Data Platform Engineer} & \parbox[c]{\hsize}{5-6} & \parbox[c]{\hsize}{500} & \parbox[c]{\hsize}{Man} & \parbox[c]{\hsize}{Australia}\\ \midrule
\parbox[c]{\hsize}{P29} & \parbox[c]{\hsize}{Data Engineer} &  \parbox[c]{\hsize}{15-20} & \parbox[c]{\hsize}{12} & \parbox[c]{\hsize}{Woman} & \parbox[c]{\hsize}{Germany}\\ \midrule
\parbox[c]{\hsize}{P31} & \parbox[c]{\hsize}{Senior SWE / Analyst} & \parbox[c]{\hsize}{200+} & \parbox[c]{\hsize}{9} & \parbox[c]{\hsize}{Man} & \parbox[c]{\hsize}{United States}\\ \midrule
\parbox[c]{\hsize}{P32} & \parbox[c]{\hsize}{SWE} & \parbox[c]{\hsize}{4} & \parbox[c]{\hsize}{20,000} & \parbox[c]{\hsize}{Woman} & \parbox[c]{\hsize}{United States}\\ \midrule
\parbox[c]{\hsize}{P34} & \parbox[c]{\hsize}{Senior SWE} & \parbox[c]{\hsize}{20} & \parbox[c]{\hsize}{50} & \parbox[c]{\hsize}{Man} & \parbox[c]{\hsize}{United States}\\ \midrule
\parbox[c]{\hsize}{P50} & \parbox[c]{\hsize}{Technical Director; Senior SWE} & \parbox[c]{\hsize}{25+} & \parbox[c]{\hsize}{10} & \parbox[c]{\hsize}{Man} & \parbox[c]{\hsize}{United Kingdom}\\ \midrule
\parbox[c]{\hsize}{P51} & \parbox[c]{\hsize}{Small Business Owner} & \parbox[c]{\hsize}{100's} & \parbox[c]{\hsize}{2} & \parbox[c]{\hsize}{Man} & \parbox[c]{\hsize}{Canada} \\ \midrule
\parbox[c]{\hsize}{P52} & \parbox[c]{\hsize}{SWE} & \parbox[c]{\hsize}{50} & \parbox[c]{\hsize}{2} & \parbox[c]{\hsize}{Man} & \parbox[c]{\hsize}{United States} \\
\bottomrule
\end{tabular}
\end{table}

\subsection{Survey}
\label{sec:survey}
We designed a 15-minute Google Forms survey on examples of implementation design decisions and considerations and distributed it to developers using the sampling strategy from Section~\ref{sec:participants}.

\subsubsection{Design}
In the survey, we first presented participants with the definition and three examples of implementation design decisions. We then asked participants to provide examples of implementation design decisions and considerations, limiting examples to the past five days in order to reduce memory bias. These questions had a response length minimum of 30 characters to encourage participants to record sufficiently detailed answers.
Following best practices, we used the HCI Guidelines for Gender Equity and Inclusivity~\cite{scheuerman2020hci} to collect gender information.
We allowed participants to select multiple responses for questions on gender. The full survey instrument is available in our supplemental materials~\cite{supplemental-materials}.

\subsubsection{Piloting}
Following best practices for experiments with human subjects in software engineering~\cite{ko2015practical}, we conducted pilots of the survey to identify and reduce confounding factors. We piloted drafts of the survey with four software developers to clarify wording and updated the survey after each round of feedback.
To ensure data quality, the survey was deployed publicly, piloted, and updated on the first 15 survey responses.

\subsection{Interviews}
\label{sec:interviews}
We conducted 45-minute interviews via online conference calls to gather examples of implementation design decisions and considerations as well as developers' decision-making process. Interviews were recorded and transcribed. Recordings were destroyed after transcription. Interview participants were compensated with a \$30 USD Amazon.com gift certificate.

\subsubsection{Design}
Topics in the interview included implementation design decisions and considerations participants made in the past five days, as well as written explicit programming strategies on how participants made their decisions. We collected written explicit programming strategies as a structured format to capture developers' processes and knowledge. An example of an explicit programming strategy from our study is in Figure~\ref{fig:ex-study-strats}; we report all collected strategies in our supplemental materials~\cite{supplemental-materials}.

Explicit programming strategies are ``human-executable procedure[s] for accomplishing a programming task"~\cite{latoza2020explicit}.
In this study, explicit programming strategies represent the process software developers used to make implementation design decisions. We collected them since developers can follow systematic processes to make some design decisions, such as selecting between multiple software architecture alternatives~\cite{bass2003software,falessi2011decision}. 
Additionally, developer expertise and processes can also be externalized by developers via explicit programming strategies~\cite{arab2022exploratory}, which allowed us to elicit software developers' decision-making processes in interviews.

\subsubsection{Protocol}
Two authors conducted the interviews: one to execute the protocol and one to record the participant's explicit programming strategy. Having a interviewer experienced in strategy authoring to write strategies ensured strategy writing quality, as authored strategies could be ambiguous or struggle to generalize~\cite{arab2022exploratory}. During the interview, we reminded participants of the definition of implementation design decisions in a Google Slides presentation, using similar wording and examples from the survey. Next, we collected implementation design decisions and considerations. Finally, for as many design decisions that time allowed for, we extracted an explicit programming strategy the participant used to make their decision. The interview protocol is available in our supplemental materials~\cite{supplemental-materials}.

To extract programming strategies, the participant explained their decision-making process step-by-step. 
Interviewers translated this to an explicit programming strategy in a shared Google Document to reduce the participant's cognitive load while recalling their process, as strategy authoring is cognitively demanding~\cite{arab2022exploratory}. The participant then reviewed the strategy and provided corrections or feedback. Because authored strategies may omit details that prevents the strategy from being usable ~\cite{arab2022exploratory}, the participants elaborated on edge cases and updated their strategy accordingly to increase its robustness. Since authored strategies' scope may be too narrow~\cite{arab2022exploratory}, interviewees updated the strategy to be general enough for a similar problem. After reviewing the strategy, the participant edited wording and clarity so the strategy was at a quality that could be released publicly. Following prior work~\cite{arab2021howtoo}, we documented a brief description of the strategy; tools, technologies, and prior knowledge necessary to use the strategy; and the steps of the strategy. The authors updated the strategy for consistency, but participants could edit the strategy upon request. Participants also could add additional data (e.g., comments, visual media), to explain their process. 

\subsubsection{Piloting}
Following best practices in experiments with human subjects in software engineering~\cite{ko2015practical}, we piloted the interview to identify and reduce confounding factors. The first author ran the interview protocol on one author and three developers and updated the protocol based on their feedback. The purpose of the pilots was two-fold: 1) to improve the clarity of the interview protocol and 2) validate the hypothesis that developers had systematic processes to make implementation design decisions and further, could articulate their processes. 
We found that all pilot participants could recall and articulate strategies for making implementation design decisions.

\begin{figure}[hbtp]
\hspace{0.1cm}

\small\textbf{Use this when:} \emph{Using less common features in libraries instead of using the popular functions}

\small\textbf{Tools/technologies:} \emph{StackOverflow, Google, continuous learning}

\small\textbf{Prior knowledge:} \emph{Common design patterns, popular libraries}

\hspace{0.1cm}

\begin{enumerate}[\itshape]
    \item\small Decide what the goal of the program is.
    \item\small Begin writing the program.
    \item\small While writing the program, search online whether other libraries support your use case. Use your prior knowledge or colleagues to help guide your search.
    \item\small Choose a library which meets your use case. This can be based on the popularity of the library with respect to the language.
    \item\small Look at the features of the library and test the ones that you’re interested in on small examples. Get a feel of the library and select a solution which achieves the desired behavior.
    \item\small If you have code that works, show the solution to another individual for review.
\end{enumerate}
\caption{An example of an explicit programming strategy from the study. \added{User-generated content is written in \emph{italics}}.
\label{fig:ex-study-strats}
}
\end{figure}

\subsection{Analysis}
\label{sec:analysis}
To analyze the collected data, we used qualitative coding. We \emph{open coded} the data for \researchquestion{1}, \researchquestion{2}, and \researchquestion{3} to summarize the data on various aspects of implementation design decisions \added{as this phenomenon is understudied}. We \emph{close coded} \researchquestion{4} to situate the strategies with prior work on software engineering expertise, which has been well-studied (e.g.,~\cite{baltes2018towards,ford2017characterizing,li2015makes,li2020distinguishes,smith2016beliefs}).

\subsubsection{Open Coding}
For \emph{open coding}, we followed best practices by Hammer and Berland~\cite{hammer2014confusing}, which outlines procedures on interpreting coding results and reporting coding disagreements. We treated generated codes as tabulated claims about the data that could be investigated in future work. We checked the reliability of the coding by resolving disagreements, first by discussing any disagreements and then coming to an agreement as a group. Finally, we interpreted coding disagreements as coding variance and reported the content of the disagreements. 

We followed best open coding practices~\cite{saldana2009coding}, preparing separate documents for each coder for qualitative analysis, taking care to remove the prompts from the responses.
In these documents, survey responses and interview responses were stored separately and analyzed independently, as the data was collected in different contexts. We also shuffled all the rows in the data to remove any ordering effects. Finally, we removed data collected from piloting. 

Open coding occurred in multiple phases. In the first phase, three authors separately reviewed the responses and inductively generated codes for each dataset. Each response was labeled with zero or more codes. Each code was given a unique identifier and a brief description. To aggregate the codes, the authors compared their separately generated codes and identified codes with similar concepts. These codes were merged under a single code and copied to a shared codebook. For the remaining codes, the authors discussed instances of disagreement and resolved them by unanimously agreeing to add or remove the code in the shared codebook. Disagreements were most frequently the result of differing scopes of codes, rather than the meaning of the participants' statements. Some disagreements arose due to an author not coding a part of the response another author did. In the second round, the authors applied the shared codebook to the original data. If there were multiple datasets to analyze for the same research question, each dataset was coded using the aforementioned process. Then, the resulting codebooks were merged by the authors. The authors identified codes with similar definitions and added them to a final codebook with a unique identifier. The remaining unmerged codes were automatically added to the final codebook. The authors performed a third round of coding with the final merged  codebook. For \researchquestion{3}, the authors then applied pattern coding to the final codes to group the codes into broader categories~\cite{miles1994qualitative}. \added{To do this, the authors placed each code into an initial category by unanimously agreeing to put it in a category or create a new one. Then, they reviewed each category and unanimously finalized its scope and, if necessary, moved the codes between categories to reflect the new scope.}

\subsubsection{Closed Coding}
For \emph{closed coding}, the first author identified a codebook to code the dataset with. For \researchquestion{4}, we used Table 3 and Table 5 from Li et al.~\cite{li2015makes}, which respectively contains codes on attributes of expert developers' decision-making processes and their software and designs. The three authors deductively applied the codes to the dataset. Each instance was labeled with zero or more codes. Next, they reconvened to discuss instances of disagreement and resolved them by unanimously agreeing on which codes were to be applied. In this step, disagreements arose due to the scope of the code rather than the meaning of the statements.

\subsubsection{Data}
For \researchquestion{1}, we analyzed 82 examples of implementation design \added{decisions from the survey and interviews}. For \researchquestion{2}, we analyzed \added{113 examples of considerations from survey and interview data as well as implementation design decisions from interviews}, since participants mentioned considerations in context of their decisions. For \researchquestion{3} and \researchquestion{4}, we analyzed a dataset with 99 steps from the 16 collected strategies.

In addition to extracting action codes for \researchquestion{3}, we analyzed the programming strategies on the decision-making process as a sequence of action codes and categories. We did this by replacing each step of the strategy with its respective code or category. If a step had multiple codes or categories, we represented the step as a sequence of codes or categories in the order they were mentioned. If a code or category occurred consecutively, they were reduced to a single occurrence. We include these representations of strategies in the supplemental materials~\cite{supplemental-materials}.

\begin{table*}\centering
\caption{The types of implementation decisions made by software developers. Codes discussed in detail are \underline{underlined}.
}\label{tab:decisions}
\begin{tabular}{p{0.47\textwidth}p{0.47\textwidth}}
\toprule
\parbox[c]{\hsize}{\textbf{Code \& Description}} & \parbox[c]{\hsize}{\textbf{Representative Quote}} \\\midrule
\parbox[c]{\hsize}{\added{\emph{Alternatives}---Deciding what high-level approaches to use to address a particular problem.}} & \parbox[c]{\hsize}{\added{``So we are ingesting data. One way...is using Python scripts... The other option that we looked into was getting it via third party tools...the third was just outsourcing it." (P28)}} \\\midrule
\parbox[c]{\hsize}{\underline{\emph{Behaviors}}---Deciding the program specification: what parameters to set for a program and their types, what outputs the program should give, and the behavior of the program.} & \parbox[c]{\hsize}{``[The API] would expect to take in the input data type, which is this union of Xarray, Numpy, Dask, all of these supported data types..." (P7)} \\\midrule
\parbox[c]{\hsize}{\emph{Data}---Deciding how to manage data within software: what data should be handled in a program, how it should be represented, and how it should be interacted with.} & \parbox[c]{\hsize}{``I opted to represent the tree as an ancestry string of the top slash the next, [and] the next. And then you can use `like' with a wild card and you'll get the subtree." (P31)} \\\midrule
\parbox[c]{\hsize}{\underline{\emph{Code constructs}}---Deciding which programming language constructs to use within a program.} & \parbox[c]{\hsize}{``Making a change to a Python program, I removed an indexing expression (\texttt{val = x[0]}) and replaced it with a destructuring assignment (\texttt{val, \_\_ = x})". (P18)}\\\midrule
\parbox[c]{\hsize}{\emph{Structure}---Deciding how to organize the codebase, where files should lie, and how code should be modularized.} & \parbox[c]{\hsize}{``I'm going to refactor this to bring out the bits of logic that pertain to...my bit of the business, so that I can then later have ownership of it...rather than [having a] big monolithic system." (P4)}\\\midrule
\parbox[c]{\hsize}{\emph{Programming languages, APIs, services}---Deciding the programming languages, APIs, or third-party services to use in the software system or script.} & \parbox[c]{\hsize}{"I used Golang to handle a large amount of JSON files that would've taken too long to handle in Python." (P20)}\\\midrule
\parbox[c]{\hsize}{\emph{Automation}---Deciding whether to implement a technology solution from scratch.} & \parbox[c]{\hsize}{``I could have manually typed in the new kinds of records that I wanted in production...but instead I encoded that all in a formalized runnable script." (P32)}\\\midrule\parbox[c]{\hsize}{\emph{Reuse}---Deciding whether code should be reused and to what extent it should be general enough to be extended to different scenarios.} & \parbox[c]{\hsize}{``Merge two C\# applications (FTP and SFTP server) into one, in order to reuse file tree state, user authentication, and so on." (P45)}\\\midrule
\parbox[c]{\hsize}{\underline{\emph{Updates}}---Deciding whether to update the software or not.} & \parbox[c]{\hsize}{``Do I tell them I can't fix the problem or do I go in and tempt small solutions, just to get it to function for a few days...or do I completely write my own fix?" (P51)}\\

\bottomrule
\end{tabular}
\end{table*}

\section{Results}
\label{sec:results}
\added{We report the results to our research questions below.} Due to space constraints, we only discuss codes we found interesting with respect to prior work. 

Prior work (e.g.,~\cite{perry1992foundations, boehm1988spiral}) characterizes implementation as naturally arising from higher levels of design. In contrast, we found that implementation design decisions involved a constant top-down and bottom-up dialogue between implementation and higher levels of design, such as requirements and architecture. \added{This supports the view that problems and software implementations co-evolve with one another~\cite{van2016decision}.} Consider Figure~\ref{fig:theory} as an example. When a developer decides whether their matrix-based data should be represented using native Python arrays, C++ arrays, or a third party library (e.g., Numpy), they may consider the architecture by thinking about what languages or third-party libraries are compatible with their system. They may also consider non-functional requirements (e.g., memory or performance) or look to how other similar modules address this problem. Finally, they may even change the priority of certain requirements, such as testability, after realizing the difficulty of prototyping unit tests with a third party library.

\subsection{What implementation design decisions do software developers make? (\researchquestion{1})}
Participants described 8 different types of implementation design decisions (see Table~\ref{tab:decisions}). 
Each code appeared more than once in our data. All codes appeared in all datasets.

\paragraph{Behaviors}
Participants decided on how the software should behave, such as its inputs, outputs, and what should occur during execution. This was an informal version of the requirements elicitation, analysis, and validation processes~\cite{paetsch2003requirements,nuseibeh2000requirements}. These decisions often required a change in requirements, \added{which corroborates the viewpoint that decisions about the solution may change requirements~\cite{van2016decision}}. For instance, participants decided on entire method specifications when requirements were under-specified and made changes to the program behavior to handle certain requirements:

\participantQuote{``But, I decided to record each line of the CSV file as a record with a header record...So on that header record, I record who gave me the file and when it was given to me. I could reproduce the CSV file from what I'm storing.}{31}

\paragraph{Code constructs}
\added{Consistent with prior work~\cite{ralph2016characteristics}}, participants debated which programming constructs to use (e.g., loop constructs, ternary operators, pointers). These decisions were the lowest level decisions made. Even at this level, participants still considered requirements (e.g., performance):

\participantQuote{[Sometimes] I would rather go for efficiency or performance [but when]..I am working with other developers, I'm more leaning to having the code more readable... Instead of functional programming mapreduce I go for loop, so that the other developers can understand the code itself. .}{22}

\paragraph{Updates}
Participants deliberated whether to make specific code changes (e.g., fixing a defect), as it could have unwanted effects. They modeled potential outcomes---a decision-making attribute of developers~\cite{li2015makes}. These decisions at times required considering the software architecture, such as while deciding whether to update dependencies:

\participantQuote{The library that we use...didn't compile anymore... We have two [options]. One is to say, `Okay, we freeze the library version...' and then we postpone the solution of the problem. Or we look at the problem and fix it immediately...}{6}

\begin{table*}\centering
\caption{Software developers' considerations for implementation decisions. Codes discussed in detail are \underline{underlined}.}\label{tab:considerations}
\begin{tabular}{p{0.47\textwidth}p{0.47\textwidth}}
\toprule
\parbox[c]{\hsize}{\textbf{Code \& Description}} & \parbox[c]{\hsize}{\textbf{Representative Quote}} \\\midrule
\parbox[c]{\hsize}{\underline{\emph{Community Support}}---How well-supported by a developer community a technology is.} & \parbox[c]{\hsize}{``I wanted to use  PHP 8.1, but there is still no general support..." (P54)}  \\\midrule
\parbox[c]{\hsize}{\emph{Features}---The features a technology contains.} & \parbox[c]{\hsize}{``Open source C\# MailKit was selected over builtin SmtpClient to...allow flexible email body manipulation." (P37)} \\\midrule
\parbox[c]{\hsize}{\emph{Popularity}---The number of users that use the technology or library.} & \parbox[c]{\hsize}{``[Laravel SPATIE Media Library] being very well understood by the rest of the Laravel developer community." (P52)}  \\\midrule
\parbox[c]{\hsize}{\emph{Reliability}---How reliable and correct the software is.} & \parbox[c]{\hsize}{``Correctness [with] concurrent updates and...mutable objects." (P44)} \\\midrule
\parbox[c]{\hsize}{\emph{Security}---How secure software is; robustness of software to adversarial attacks.} & \parbox[c]{\hsize}{``Each decision had tradeoffs...[in] security (the latter being exposing, possibly private, brands.)" (P24)} \\\midrule
\parbox[c]{\hsize}{\emph{Maintainability}---How easily maintenance actions (e.g., fixing defects, updating components) can be performed on software.} & \parbox[c]{\hsize}{``Over-engineering a system that may...add cognitive overhead to any maintenance." (P39)} \\\midrule
\parbox[c]{\hsize}{\emph{Testability}---How easily software can be tested (e.g., unit tests).} & \parbox[c]{\hsize}{``Testability (functional is almost always easier to test)." (P27)} \\\midrule
\parbox[c]{\hsize}{\emph{Extensibility}---How easily the code can be extended to accommodate changes (e.g., new features).} & \parbox[c]{\hsize}{``So how the structure and application itself is laid out so that it's not going to be a pain to work with, as we expand it." (P50)} \\\midrule
\parbox[c]{\hsize}{\emph{Performance}---Performance aspects of the code (e.g., runtime, memory).} & \parbox[c]{\hsize}{``The function call is expensive in certain...circumstances." (P18)} \\\midrule
\parbox[c]{\hsize}{\emph{Reproducibility}---Whether code is able to reproduce the same output, given the same input.} & \parbox[c]{\hsize}{``Does the code do it in an idempotent way? So...the same input would do the same output regardless of how many times you do it." (P22)} \\\midrule
\parbox[c]{\hsize}{\emph{Requirements}---The requirements of the software; customer needs.} & \parbox[c]{\hsize}{``After the first implementation, a new requirement came so structuring for reuse [was] useful. " (P15)} \\\midrule
\parbox[c]{\hsize}{\underline{\emph{Future Requirements}}---Requirements or customer needs that may or may not occur in the future.} & \parbox[c]{\hsize}{``Will there be a need to run the pipeline in parallel some day (like on Spark or Dask)?" (P27)} \\
\midrule
\parbox[c]{\hsize}{\emph{Skills}---The current skills of the team or of the developer.} & \parbox[c]{\hsize}{``We also chose PHP because...more developers familiar with the PHP framework than with Python frameworks." (P56)}\\
\midrule
\parbox[c]{\hsize}{\emph{Budget}---Amount of resources (e.g., time, money) available to implement the software project.} & \parbox[c]{\hsize}{``Because we were on a tight deadline...I decided to just process [the data] all on my local machine...and then upload it." (P57)}\\
\midrule
\parbox[c]{\hsize}{\emph{Reusing Resources}---Reusing existing resources (e.g., code, practices).} & \parbox[c]{\hsize}{``What parts of the code will they reuse?" (P32)}\\
\midrule
\parbox[c]{\hsize}{\emph{Difficulty}---How much effort completing the implementation will be.} & \parbox[c]{\hsize}{``The implementation difficulty comes into [these decisions]." (P28)}\\
\midrule
\parbox[c]{\hsize}{\emph{Readability}---How easily code syntax is read by a developer.} & \parbox[c]{\hsize}{``Generics in code may be harder to comprehend for some..."(P42)}\\\midrule
\parbox[c]{\hsize}{\emph{Code Cleanliness}---The quality of the implementation's code; how easy it is to onboard other developers and make updates.} & \parbox[c]{\hsize}{``Try not to make ravioli code where we have too many modules that do little things." (P42)}\\
\midrule
\parbox[c]{\hsize}{\emph{Simplicity}--The length or complexity of the implementation.} & \parbox[c]{\hsize}{``I did this to keep my pull request shorter and closer to the original code. Less to read means faster code review." (P18)}\\\midrule
\parbox[c]{\hsize}{\underline{\emph{Consistency}}---Being consistent with the code style of the programming language or code base.} & \parbox[c]{\hsize}{``So, not only is it existing code that's already there. I don't want to be the person to introduce something different." (P34)}\\
\midrule
\parbox[c]{\hsize}{\underline{\emph{System Fit}}---How well the implementation fits in with an existing code base or system.} & \parbox[c]{\hsize}{``Where to set up the event subscription... [In] a React component or outside of the React/Redux content..." (P24)}\\
\midrule
\parbox[c]{\hsize}{\emph{Data}---How data in the system will be managed or handled.} & \parbox[c]{\hsize}{``This is a trade-off of having less-fresh data, with being more robust in the event the 3rd party is unavailable." (P31)}\\
\midrule
\parbox[c]{\hsize}{\emph{Impacts}---The impacts that the implementation may cause.} & \parbox[c]{\hsize}{``I want to be very safe when making potentially-impactful changes in production environments." (P59)}\\
\midrule
\parbox[c]{\hsize}{\emph{Users}---Thinking about collaborators who will be working in the code base; the usability of the software for end-users.} & \parbox[c]{\hsize}{``It's trying to make the code...understandable for the other developers for maintenance purposes and if they need to upgrade the code..." (P22)}\\
\midrule
\parbox[c]{\hsize}{\emph{Documentation}---Writing documentation for the implementation.} & \parbox[c]{\hsize}{``So that’s a significant documentation...cost." (P50)}\\
\bottomrule
\end{tabular}
\end{table*}
\subsection{What considerations do software developers have while making implementation design decisions? (\researchquestion{2})}
Participants described 25 distinct considerations while making implementation design decisions. We report them in Table~\ref{tab:considerations}. Each code appeared more than once---17 codes appeared in all datasets, 7 codes appeared in two datasets, and 1 code appeared in one dataset.

\paragraph{Community support}
Participants reported that community support for third-party libraries was a factor. This ensured that dependencies were well-maintained for software quality. Having community support enables the production of educational resources for the tool, such as on YouTube~\cite{macleod2015code} as well as StackOverflow and blog posts~\cite{parnin2012crowd}. Participants said this reduced onboarding costs:

\participantQuote{And there are instructional videos on YouTube and whatnot [that] can already teach people how to do [use the tool] without the rest of the development team having to do anything.}{52}

\paragraph{Future requirements}
Similar to the management phase in requirements engineering~\cite{paetsch2003requirements,nuseibeh2000requirements} participants noted requirements could change. 
Understanding future requirements ensured the software was useful in the long term. Estimating them depended on prior experience and domain knowledge:

\participantQuote{...I'm making an assumption about what might come down the pipe in the future. [It] depends on kind of my experience in that field, and my work that I've done with past clients to think that my future clients might be similar enough to them.}{51}

\paragraph{Consistency}
\added{Similar to prior work~\cite{ralph2016characteristics}}, consistency of the code style in the codebase or following programming language convention was a consideration. This reduced confusion between teammates and cognitive overhead for individuals working across multiple contexts. This occurred both at the application- and module-levels:

\participantQuotes{I have 200 or so web applications and having consistency...makes it easy for me to switch between them without having to re-remember a whole different framework.}{31}{So some places [a stock] is called a stock, some places it's called security...I will try to keep that pattern, even if it's something I don't necessarily agree with.}{34}

\paragraph{System fit}
Participants considered how easily the implementation could be integrated with existing code, such as synergy with specific technologies. This is an attribute of expert developers' software and designs~\cite{li2015makes}. System fit required knowledge of the system's architecture:

\participantQuote{Choosing between django-q and celery was difficult- one is closely coupled with django environment and the other has long history/reliability.}{55}

\subsection{What process do software developers follow to make implementation design decisions? (\researchquestion{3})}
\begin{table*}\centering
\caption{Software developers' actions in making implementation decisions. Codes discussed in detail are \underline{underlined}. The number of times an action category is repeated within the same strategy in our data is denoted with $\times$.}\label{tab:actions}
\begin{tabular}{p{0.47\textwidth}p{0.47\textwidth}}
\toprule
\parbox[c]{\hsize}{\textbf{Code \& Description}} & \parbox[c]{\hsize}{\textbf{Representative Quote}}\\\midrule
\multicolumn{2}{c}{\parbox[c]{0.96\textwidth}{\textbf{Defining Problem Space ($\times8$)}}}\\\midrule
\parbox[c]{\hsize}{\emph{Providing Context}---Explaining context about the problem the developer is facing (e.g., refactoring).} & \parbox[c]{\hsize}{``Write an initial program...If you have another program that requires a similar behavior, consider whether you want to refactor the code." (P15)} \\\midrule
\parbox[c]{\hsize}{\emph{Defining Requirements}---Defining the requirements of the solution, considering user needs, business needs, and organization needs.} & \parbox[c]{\hsize}{``...Figure out what use cases [your end users] would want for this function. Ask your end users to provide examples of inputs..." (P7)} \\\midrule
\parbox[c]{\hsize}{\underline{\emph{Updating Requirements}}---Updating the requirements of the solution after they are initially defined.} & \parbox[c]{\hsize}{``If you find new requirements from your proof-of-concept, go to step 1." (P52)}\\\midrule
\parbox[c]{\hsize}{\textbf{Ideating} ($\times0$)}\\\midrule
\parbox[c]{\hsize}{\emph{Brainstorming}---Brainstorming potential solutions that could solve the problem.} & \parbox[c]{\hsize}{``Think about the problem for a set period of time and write down more than one idea on how to implement a solution..." (P7)}\\\midrule
\parbox[c]{\hsize}{\textbf{Assessing} ($\times4$)}\\\midrule
\parbox[c]{\hsize}{\underline{\emph{Evaluating}}---Evaluating the developer's current situation; considering the pros and cons for each solution.} & \parbox[c]{\hsize}{``List out all the options that you have into a document and their pros and cons." (P29)}\\\midrule
\parbox[c]{\hsize}{\emph{Estimating}---Estimating the potential costs associated with implementation.} & \parbox[c]{\hsize}{``Determine whether the requirements are realistic given the resources you have available." (P34)}\\\midrule
\parbox[c]{\hsize}{\textbf{Prototyping} ($\times1$)}\\\midrule
\parbox[c]{\hsize}{\underline{\emph{Proof-of-Concept}}---Building a proof-of-concept for a potential solution.} & \parbox[c]{\hsize}{``Test each option in the development environment..." (P29) }\\\midrule
\parbox[c]{\hsize}{\textbf{Implementing} ($\times15$)}\\\midrule
\parbox[c]{\hsize}{\emph{Choosing}---Choosing a solution to implement.} & \parbox[c]{\hsize}{``Select the option which meets your requirements..." (P18)}\\\midrule
\parbox[c]{\hsize}{\emph{Planning}---Planning the steps needed to implement the solution.} & \parbox[c]{\hsize}{``List out the tasks that need to be done based on the requirements of the problem/client and the technology stack available." (P51)}\\\midrule
\parbox[c]{\hsize}{\emph{Implementing}---Implementing a particular solution.} & \parbox[c]{\hsize}{``Build an implementation from the proof of concept." (P52)}\\\midrule
\parbox[c]{\hsize}{\emph{Updating Implementation}---Trying a new implementation or updating an existing one based on previous implementation attempts.} & \parbox[c]{\hsize}{``If there is a problem with the solution that’s implemented, go to step 1 with what you learned by implementing the solution." (P29)}\\\midrule
\parbox[c]{\hsize}{\emph{Deploying}---Releasing the solution to the public.} & \parbox[c]{\hsize}{``Implement your solution...and deploy it into a development environment." (P28)}\\\midrule
\parbox[c]{\hsize}{\textbf{Verifying} ($\times1$)}\\\midrule
\parbox[c]{\hsize}{\emph{Reviewing}---Having others review and provide feedback to the solution.} & \parbox[c]{\hsize}{``Have others review your implementation proposal (over coffee can help)." (P28)}\\\midrule
\parbox[c]{\hsize}{\emph{Testing}---Testing the implementation for functionality and potential defects.} & \parbox[c]{\hsize}{``Test your implementation until you find most of the bugs you can and your teams agree to release to prod." (P52)}\\\midrule
\multicolumn{2}{c}{\parbox[c]{0.96\textwidth}{\textbf{Updating Knowledge} ($\times4$)}}\\\midrule
\parbox[c]{\hsize}{\underline{\emph{Researching}}---Learning more about the %
problem %
or potential solutions.} & \parbox[c]{\hsize}{``Search online whether other libraries support your use case." (P6)}\\
\bottomrule
\end{tabular}
\end{table*}

We describe the types of actions developers take while making implementation design decisions (Sections~\ref{sec:actions}). We then report the \added{sequence of actions that developers follow in their decision-making process} (Section~\ref{sec:sequences}).

\subsubsection{Actions}
\label{sec:actions}
Participants described 15 types of actions in their strategies. We grouped these into 7 categories: defining the problem space; ideating, evaluating, prototyping, implementing, and verifying potential solutions; and updating knowledge. Many of these actions overlapped with the traditional software development process, similar to prior work~\cite{sharif2013identifying}. Furthermore, participants' actions were often a dialogue between the implementation and requirements. The full list of the actions in implementation design decisions are in Table~\ref{tab:actions}. All codes occur in the data at least twice.

\paragraph{Updating requirements}
Study participants described times when requirements changed after they were defined. This occurred after learning from proof-of-concepts, reacting to changes in the situation, or analyzing the requirements' viability. Unlike in requirements engineering~\cite{paetsch2003requirements,nuseibeh2000requirements}, requirements also changed during implementation. In these cases, updating requirements was how participants dealt with unforseen circumstances during implementation, such as time constraints. Participants even developed heuristics to do so: 

\participantQuote{If you are under time constraints, restrict the scope of your implementation and don’t let perfect be the enemy of good.}{50}

\paragraph{Evaluating}
Participants evaluated potential alternatives for pros and cons, where they compared them against a list of considerations, especially non-functional requirements. Some participants wrote lists or drew matricies, while others developed checklists from their expertise:

\parbox{0.95\linewidth}{
    \vspace{5pt}
    \small
    \faQuoteLeft\xspace
    \emph{Decide what you think is a pro/con of a given solution for your use case. 
        \begin{itemize}
            \item Security concerns, an insecure package is never acceptable... 
            \item The popularity of the package is critical for evaluating lifetime reliability and long term support. 
            \item Level of skill required to use the package, poorly designed apis will be difficult to extend if needs change, and complicated for junior developers to work with. 
            \item Clean, consistent and clear are the ideal interfaces. 
            \item Consider the cost to replace the package if licensing, support or project direction dramatically change...\normalfont{" (P50)}
        \end{itemize}
    }
    \vspace{5pt}
}

\paragraph{Proof-of-concept}
Study participants reported creating proof-of-concepts, which is also important in requirements elicitation~\cite{paetsch2003requirements,nuseibeh2000requirements}. This quickly determined whether a potential solution was viable. Participants varied in which solutions they chose to prototype---some chose to prototype only the best candidate solution, while others prototyped all potential solutions. Prototyping was also an information gathering mechanism to brainstorm potential solutions:

\participantQuote{Look at the features of the library and test the ones that you're interested in on small examples. Get a feel of the library...}{6}

Participants developed a proof-of-concept to update requirements. This was one way they considered a higher level of design during implementation design decisions:

\participantQuote{Ask the people who you interviewed to try your function and provide feedback on any of the parameters.}{7}

\paragraph{Researching}
Participants researched the problem space, \added{as identified in prior work~\cite{ralph2016characteristics}}. This was the second most reported action. This action was often used to elicit requirements. Participants worked with stakeholders (e.g., project managers) and accessed websites for knowledge sharing in software engineering, such as StackOverflow~\cite{parnin2012crowd} and Reddit~\cite{hardin2016learning}. They reviewed similar projects and used empirical methods to understand the problem and generate requirements:

\participantQuote{[Using] tools to go through Git history...to find potential problems...}{4}

\begin{table}\centering
\caption{Median position of actions in participants' strategies and percent of strategies containing actions.}\label{tab:actions-positions}
\begin{tabular}{p{0.33\linewidth}p{0.25\linewidth}p{0.25\linewidth}}
\toprule
\multirow{2}{*}{\textbf{Action}} & \multirow{2}{*}{\textbf{Median Position}} & \parbox[c]{\hsize}{\hfil \textbf{\% Strategies w/}} \\
& & \hfil \textbf{Action} \\\midrule

Providing Context & \hfil 1.5 & \hfil 12.5\% \\\midrule
Researching & \hfil 2 & \hfil 75.0\% \\\midrule
Defining Requirements & \hfil 2 & \hfil 81.3\% \\\midrule
Brainstorming & \hfil 3 & \hfil 62.5\% \\\midrule
Estimating & \hfil 3 & \hfil 25.0\%\\\midrule
Evaluating & \hfil 4 & \hfil 81.3\%\\\midrule
Choosing & \hfil 5 & \hfil 81.3\%\\\midrule
Planning & \hfil 6 & \hfil 25.0\%\\\midrule
Proof-of-Concept & \hfil 6.5 & \hfil 31.3\%\\\midrule
Updating Requirements & \hfil 7 & \hfil 43.8\%\\\midrule
Implementing & \hfil 7 & \hfil 68.8\%\\\midrule
Reviewing & \hfil 8 & \hfil 43.8\% \\\midrule
Testing & \hfil 9  & \hfil 31.3\%\\\midrule
Updating Implementation & \hfil 9 & \hfil 18.8\%\\\midrule
Deploying & \hfil 10.5  & \hfil 12.5\%\\\bottomrule

\end{tabular}
\end{table}

\subsubsection{Processes}
\label{sec:sequences}
Just as how software designers follow individualized processes~\cite{sharif2013identifying}, we found that strategies \added{about developers' decision-making processes} were unique: there were no repeated sequences of action codes or categories \added{in the strategies}. One source of dissimilarity were action codes and categories that were repeated in strategies. Table~\ref{tab:actions} shows the occurrences of repeated action categories in the strategies. All action categories were repeated except \code{ideating} actions. \code{Implementing} and \code{defining problem space} actions were most repeated in participants' strategies. Yet, there were commonalities in the structure, namely when certain types of actions occur. This is shown in Table~\ref{tab:actions-positions}, which reports the median position of each action category across all action sequences.

\begin{table}\centering
\caption{Frequency of developer expertise 
from Li et al.~\cite{li2015makes} in participants' strategies.}\label{tab:expertise}
\begin{tabular}{p{0.8\linewidth}p{0.08\linewidth}}
\toprule
\textbf{Expertise (quoted from Li et al.~\cite{li2015makes})} & \textbf{Count}\\\midrule
\textbf{Decision Making} & \\\midrule
Knowledgeable about customers and business & \hfil 56 \\\midrule

Sees the forest and the trees & \hfil 45 \\\midrule

Knowledgeable about tools and building materials & \hfil 39 \\\midrule

Knowledgeable about their technical domain & \hfil 38 \\\midrule

Knowledgeable about engineering processes & \hfil 31\\\midrule

Models states and outcomes & \hfil 24 \\\midrule

Handles complexity & \hfil 24 \\\midrule

Knowledgeable about people and the organization & \hfil 13 \\\midrule

Updates their mental models & \hfil 8 \\\midrule

\textbf{Software \& Designs} & \\\midrule

Carefully constructed & \hfil 26 \\\midrule

Fitted & \hfil 11 \\\midrule

Evolving & \hfil 10 \\\midrule

Attentive to details & \hfil 9 \\\midrule

Anticipates needs & \hfil 5 \\\midrule

Creative & \hfil 5 \\\midrule

Long-termed & \hfil 2 \\\midrule

Elegant & \hfil 2 \\\midrule
\end{tabular}
\end{table}

\subsection{Which types of developer expertise are described in the implementation design decision-making process? (\researchquestion{4})}

Table~\ref{tab:expertise} shows the types of developer expertise in study participants' decision-making process. We found that decision-making expertise was most commonly cited in strategies. Expertise relating to deeply understanding the vision of the project (e.g., \code{knowledgeable about customers and business})
was most frequently referenced. Expertise on evaluating the pros and cons of a solution (e.g., \code{makes tradeoffs})
was also mentioned. Participants also frequently described forms of technical expertise (e.g., \code{knowledgeable about tools and building materials}).

Expertise that was less commonly cited in strategies largely related to aspects of the software and designs (e.g., \code{attentive to details}).
Expertise about teammates or the company (e.g., \code{knowledgeable about the people and organization}) and responding to changing problem contexts (e.g., \code{updates their mental models}) were also less referenced. 

\section{Threats to validity}
\label{sec:threats-to-validity}

\subsubsection{Internal validity}
Strategy content may have been influenced by the study authors since they were initially recorded by them. To address this, the authors asked participants to review and confirm the strategy multiple times to ensure the strategy was accurate and written as the participants wished. Participants could also directly make changes to the strategy upon request.

\added{The authors could have confirmation biases that developer actions must follow normative theories of the software development life cycle, which could influence the generation of action codes. We reduced this threat by achieving consensus on each code applied in our qualitative analysis. Future studies using other methods, such as contextual inquiries of implementation design decision-making, could address this bias.}

Additionally, memory bias could have introduced inaccuracies in participants' recall on past decisions, considerations, and strategies. We reduced this bias by asking participants to recount decisions, considerations, and actions that occurred in the past five days in both the survey and interview.

Participants could have been primed in the implementation design decisions they reported from the survey and interview examples. We reduced this threat by providing a short, diverse set of examples to show the breadth of the phenomenon.

Study participants may have misunderstood the wording of the questions. To reduce this threat, we piloted the survey and interview with software developers and study team members and asked for feedback on clarity. We also performed pilots on the first 15 survey responses for data quality. 

\subsubsection{External validity}
Any small-scale empirical study has generalizability issues~\cite{flyvbjerg2006five}. To address this, we sampled Reddit users across a diverse set of subreddits. Our sample represents diverse geographic regions, engineering organization sizes, roles, and amounts of relevant experience. Additionally, we also collected the data on decisions and considerations using two methods, which we used to corroborate results.

Empirical studies can also suffer from selection bias. The subreddits we recruited from may be homogeneous due having to common interests, so some programming expertise was not represented in our study (e.g., game development). Our survey also limited representation to regions where English is a primary language due to the survey being written in English. We addressed this threat by ensuring the survey was as short as possible, accurately advertising the survey's length, and providing incentives for participating in the interview.

One common issue for programming strategies is that they may not generalize due to defects or having a narrow scope~\cite{arab2022exploratory}. To address this concern, we asked participants to test and fix their strategy and verify whether their strategy was generalized. The study authors were familiar with writing high-quality and generalized strategies.

\subsubsection{Construct validity}
Since the participants' decisions, considerations, and strategies were self-reported, there could be inconsistencies in what participants report doing versus what they actually did. They may have forgotten to explicitly mention actions or misremembered the process they used.

\section{Discussion \& Future Work}
\label{sec:discussion}

Our findings overlap some with prior work (e.g.,~\cite{li2015makes,petre2016software,balaji2012waterfall, petre2019software,ralph2016characteristics}). In this section, we discuss them in relation to implementation design decisions. This produces several implications, which we elaborate on.

Our results suggest that implementation design decisions are shaped by higher levels of design (e.g., requirements and architecture). Also, a developer's decisions can directly and intentionally shape these higher level concerns. Thus, interpreting requirements throughout implementation is key to making these decisions. Requirements appeared in the decisions (e.g., \code{behaviors}), considerations (e.g., \code{future requirements}), and actions (e.g., \code{defining requirements}) of software developers. It also was the most frequently cited form of expertise (e.g., \code{knowledgeable about customers and business}) in strategies. This highlights the perspective that requirements engineering is an ongoing process throughout implementation and maintenance. Depending on how much control the developer had, they could re-interpret or completely change requirements, suggesting that understanding \emph{how} to update requirements to match dynamic contexts could be a software engineering skill. This contrasts the notion of requirements being set prior to implementation. Rather, it supports prior work stating that requirements can be iterated upon through prototyping~\cite{paetsch2003requirements,nuseibeh2000requirements,boehm1988spiral}.

Next, we find that maintainability is a major theme in implementation design decisions. It appeared in the decisions (e.g., \code{reuse}) and considerations (e.g., \code{extensibility}, \code{reusing resources}) as it reduced workload, cognitive load, and technical debt. This suggests that software developers may need to develop a sense of how to anticipate maintenance effort and develop knowledge on how to manage and reduce debt, such as separation of concerns and modularity.

We also find that the process to make implementation design decisions is both an art and a science. Across different developers and problems, there were strong commonalities in the considerations and strategy structure. However, each strategy was unique---prior work has shown that software development teams also follow their own individual processes for early stage software design work~\cite{sharif2013identifying}. This implies that making implementation design decisions has a common structure. Yet, it requires expert judgment developed from experience with similar problems to know when to deviate from it for the given use case. Furthermore, it suggests this form of design expertise is both systematic and opportunistic, which has been observed in prior work~\cite{davies1991characterizing, zannier2007model}. One source of opportunism is when similar types of actions are repeated in different parts of the strategy. This suggests that implementation design decisions require rounds of iteration in different stages, especially in \code{implementing} and \code{defining the problem space}.

Finally, our results suggest that some forms of expertise are more implicit, while others are more tacit. Explicit programming strategies are a form of explicit programming knowledge~\cite{latoza2020explicit}. Participants' strategies largely referenced decision-making expertise, implying that it could be explicit knowledge. Meanwhile, expertise on developing software and designs was referenced noticeably less frequently in participants' strategies but instead overlapped with our enumerated considerations. This suggests that this form of knowledge is more tacit and is applied in the moment of programming problem-solving. 

These implications affect software engineering researchers, educators, and practitioners. We describe how our findings apply to them and provide opportunities for future work.

\subsection{Educators}
Our findings have implications on how to teach programming and software design. While teaching software engineering, educators could consider providing open-ended projects, as suggested by Offutt and Baral~\cite{offutt2022designing}. This would provide students with opportunities to make various types of implementation design decisions. Further, these projects could span for a longer duration to teach students how their decisions shape software maintenance.
Educators could scaffold students' problem-solving process by authoring their own explicit programming strategies on how to make implementation design decisions. 
Educators could also use the list of considerations from Table~\ref{tab:considerations} as a checklist for students to follow when evaluating candidate solutions. 

\subsection{Software Engineers}
Our findings can help novice engineers to make better implementation design decisions. Since iteration is an important part of making implementation design decisions, less experienced engineers could work prototyping into their regular practice and become accustomed to learning from small-scale failures through prototyping. Managers or mentors could reinforce this learning environment by encouraging this practice. 

Additionally, novices could also use the list of considerations as a checklist to help make decisions. More senior engineers could extend our list of considerations by writing their own definitions and heuristics. They could also add their own considerations to teach less experienced team members.

\subsection{Researchers}
This work raises questions about our understanding of software design. Previous work viewed software design as a sociotechnical process (e.g.,~\cite{mangano2014supporting}), a set of habits (e.g.,~\cite{petre2016software}), or as high-level code structure (e.g.,~\cite{perry1992foundations}). \added{Our work extends this knowledge by focusing on the cognitive process involved in software design. We view software design as a decision-making exercise,} following prior work (e.g.,~\cite{ralph2016characteristics, van2016decision}).

There are several directions that require additional study. Future work could study each of the the consideration codes to discover how developers estimate them and how they compare considerations against one another. This could help develop automated metrics or tools to aid developers' decision-making. Additionally, future work could examine how experts' decision-making processes differ than that of novices', especially in the strategy structure and considerations. This could help better understand the attributes that relate to effective decision-making processes and advance understanding of software engineering expertise.

\section*{Data Availability}
Our supplemental materials are available on Figshare~\cite{supplemental-materials}. Data includes the codebooks for \researchquestion{1}, \researchquestion{2}, and \researchquestion{3}; the plain text, action code, and action category representations of participants' strategies; the survey instrument; and the interview protocol.

\section*{Acknowledgments}
\added{
We thank our survey and interview participants for their insight and Soham Pardeshi, Lilian Liang, Nimit Johri, and Tobias D{\"u}rschmid for their feedback. We give special thanks to Mei \meiicon, an outstanding canine software engineering researcher, for providing support and motivation throughout this study.

This work was supported by the National Science Foundation under grants 1539179, 1703734, 1703304, 1836813, 1845508, 2031265, 2100296, 2122950, 2137834, 2137312, and by unrestricted gifts from Microsoft, Adobe, and Google.
}

\bibliographystyle{IEEEtran}
\bibliography{bibliography}
\end{document}